\documentclass[12pt]{article}
\usepackage{graphicx}

\usepackage{amsmath}
\usepackage{amssymb}

\newcommand*\icarus{Icarus}
\newcommand*\mnras{MNRAS}

\newcommand*\physrep{Phys. Rep.}

\newcommand*\prl{Phys. Rev. Lett.}
\newcommand*\pra{Phys. Rev. A}
\newcommand*\pre{Phys. Rev. E}

\begin{document}

\title{\bf Dynamical entropy \\ of the separatrix map}

\author{Ivan~I.~Shevchenko\/\thanks{E-mail:~ivan.i.shevchenko@gmail.com} \\
Saint~Petersburg State University, 7/9 Universitetskaya nab., \\
199034 Saint~Petersburg, Russia  \\
Institute of Applied Astronomy of the Russian Academy of Sciences, \\
191187 Saint~Petersburg, Russia}

\date{}

%\date{\today}

\maketitle

\begin{center}
Abstract
\end{center}

\noindent
We calculate the maximum Lyapunov exponent of the motion in the separatrix map's
chaotic layer, along with calculation of its width, as functions of the adiabaticity parameter
$\lambda$. The separatrix map is set in natural variables; and the case of the layer's
least perturbed border is considered, i.~e., the winding number of the layer's border (the
last invariant curve) is the golden mean. Although these two dependences (for the
Lyapunov exponent and the layer width) are strongly non-monotonous and evade any
simple analytical description, the calculated dynamical entropy $h$ turns out to be a
close-to-linear function of $\lambda$. In other words, if normalized by $\lambda$, the
entropy is a quasi-constant. We discuss whether the function $h(\lambda)$ can be in fact
exactly linear, $h \propto \lambda$. The function $h(\lambda)$ forms a basis for
calculating the dynamical entropy for any perturbed nonlinear resonance in the first
fundamental model, as soon as the corresponding Melnikov--Arnold integral is estimated.

\newpage

\section{Introduction}
\label{intro}

In \cite[p.~37]{C71CERN}, Boris Chirikov, when considering the
Krylov--Kolmogorov (dynamical) entropy, notes that the term
entropy here ``cannot be regarded as felicitous, because there is
confusion with the usual thermodynamic entropy. In fact these
quantities are completely different even in dimension.'' Then, by
introducing a minimum permissible volume of phase space, defined
as the correlation volume (inside which the motion is not
chaotic), Chirikov arrives to a dynamical characterization of
entropy of a chaotic system. Indeed, the correlation volume
exponentially decreases with time, opposite to the usual
definition case for the minimum volume, when the latter is
determined by the phase space quantization and is constant in
time. Then, ``defined, in such a way, the dynamical entropy
permanently increases with time (in a state of statistical
equilibrium!) for any system with mixing''~\cite[p.~38]{C71CERN}.
The following formula for the dynamical entropy naturally arises:

\begin{equation}
h = \lim_{t \to \infty} \left( - \frac{1}{t} \int \ln \Delta
\mu_\mathrm{c}(t) d \mu \right) , \label{eq_de1}
\end{equation}

\noindent where $\mu$ is the phase space measure,
$\mu_\mathrm{c}(t)$ is the correlation volume as a function of
time $t$, and the integration is performed over the whole chaotic
domain where the motion takes place; see
\cite[eq.~(2.3.14)]{C71CERN}. The dynamical entropy $h$ has
dimension of frequency. In \cite{C71CERN,BGS76PRA}, it is called
K-entropy (Kolmogorov entropy); and, in \cite{C79PhR}, KS-entropy
(Kolmogorov--Sinai entropy).

On assuming that the local Lyapunov exponent value is constant
over the chaotic domain, formula~(\ref{eq_de1}) is
straightforwardly reducible to an approximate one, on other
grounds proposed in \cite[eq.~(6)]{BGS76PRA}:

\begin{equation}
h = L \mu_\mathrm{ch} ,
\label{eq_de2}
\end{equation}

\noindent where $L$ is the maximum Lyapunov exponent in the
chaotic domain, and $\mu_\mathrm{ch}$ is the latter's measure.

On variations of any parameter of a system under study, appearance/dis\-appearance of
regular islands (including those due to marginal resonances) in the phase space often
induce disturbances in the dependences of $L$ and $\mu_\mathrm{ch}$ on the parameter.
However, in the dependence of the product $L \mu_\mathrm{ch}$ on the given parameter
these disturbances are usually observed to compensate each other, so that the resulting
dependence is smooth; spectacular examples can be found in \cite[figs.~1--4]{S04PLA} for
the standard map case; and, for the separatrix map, in \cite[figs.~1b, 1c]{S08MN}. This
smoothing-out effect graphically demonstrates, as nothing else, the fundamental character
of the dynamical entropy.

Generally, the separatrix maps describe the motion in the vicinity
of the separatrices of nonlinear resonances subject to periodic
perturbations. Construction and analysis of separatrix maps
constitute a powerful tool of modern nonlinear dynamics
\cite{C79PhR,AKK96PhyD,AZ95PhPl,AZ96PhPl,V96JETP,V99JETP,S99CM,S00JETP,A06LNP,A14,S20ASSL}.

Under general conditions~\cite{C79PhR,LL92}, a model of a
nonlinear resonance is provided by the nonlinear pendulum with
periodic perturbations. This is the so-called first fundamental
model of perturbed nonlinear resonance, see a review in
\cite{S24JETPL}. Its Hamiltonian is given by

\begin{equation}
H = {{{\cal G} p^2} \over 2} - {\cal F} \cos \varphi + a \left(
\cos(k \varphi - \tau) + \cos(k \varphi + \tau) \right), \label{h}
\end{equation}

\noindent where $\tau = \Omega t + \tau_0$. The first two terms in
Eq.~(\ref{h}) represent the Hamiltonian $H_0$ of the unperturbed
pendulum, while the two remaining ones the periodic perturbations.
The variable $\varphi$ is the resonance phase; $\tau$ is the phase
angle of perturbation; $\Omega$ is the perturbation frequency, and
$\tau_0$ is the initial phase of the perturbation; $p$ is the
momentum; ${\cal F}$, ${\cal G}$, $a$, and integer or half-integer
$k$ are constant parameters.

A number of problems on non-linear resonances in mechanics and
physics is described by the perturbed pendulum-like
Hamiltonian~(\ref{h}). The case of symmetric perturbation $a = b$
with $k = 1$ is of especial interest. In this case, the
Hamiltonian~(\ref{h}) describes, in particular, the pendulum with
the vertically oscillating point of suspension. The
near-separatrix motion of system~(\ref{h}) in this case was
demonstrated in \cite{C79PhR,S00JETP} to be effectively described
by the classical separatrix map.

Models with arbitrary non-zero values of integer $k$ and zero
either $a$ or $b$ concern the problem of a particle motion in the
field of two planar waves \cite{ED81JSP,E85PhR,ZA95PRE}. In
celestial mechanics, models with $k = 1/2$ and specific $a$ and
$b$ values were applied to describe dynamics in vicinities of the
3/1 orbital resonance in planetary satellite systems
\cite{Ma90Ic,S20ASSL}.
A model with $k = 1$ and $b = - a/7$
describes rotational dynamics close to synchronous spin-orbit
resonance, of non-spherical satellites in elliptic orbits
\cite{WPM84Ic,Ce90ZAMP}. By introducing {\it separatrix
algorithmic maps} \cite{S99CM,S20ASSL}, the separatrix map theory
can be applied to a broad spectrum of physical phenomena
(including the just mentioned ones), described by
Hamiltonian~(\ref{h}).

Quantized versions of separatrix maps are broadly used in studies
of quantum chaos, see \cite{Q24x} and references therein.

Therefore, the employed model relates with particularly
interesting physical phenomena (such as the pendulum with the
vibrating suspension point) and is both applicable to a broad
spectrum of physical phenomena.

However, little is known nowadays on the Lyapunov exponents and
the dynamical entropy of the separatrix maps, including its
classical version. To elucidate the properties of the dynamical
entropy of the classical separatrix map is the major goal of the
present study.

We calculate the maximum Lyapunov exponent of the motion in the
separatrix map's chaotic layer, along with calculation of its
width, as functions of the adiabaticity parameter. The
adiabaticity parameter $\lambda$ is defined as the ratio of the
perturbation frequency $\Omega$ to the frequency $\omega_0$ of
small-amplitude oscillations on resonance
\cite{C79PhR,S00JETP}.\footnote{The {\it adiabaticity parameter}
$\lambda$ should not be confused with the {\it adiabatic
invariant}, which is a completely different notion, whose theory
is considered e.g. in \cite{LL92}.}

We derive a simple approximate formula for the dynamical
entropy. This might be useful in many applied problems. Indeed,
computations of the Lyapunov exponents and, especially, measure of
chaotic domains often represent complicated numerical tasks.
Therefore, as soon as a formula connecting these two quantities is
obtained, any computation of the first one automatically provides
the value of the second one, and vice versa. This is especially
useful for estimations of chaotic domains' measure, which is
often much more complicated to obtain in numerical experiments, in
comparison with Lyapunov exponents. What is more, this may provide
insights for theoretical studies, which could provide analytical
explanations for the observed relationships.

\section{Separatrix map}
\label{sec_sxmap}

Chirikov~\cite{C79PhR} derived the separatrix map that describes
the motion in the vicinity of the separatrices of
Hamiltonian~(\ref{h}) with $k=1$:

\vspace{-3mm}

\begin{eqnarray}
& & w' = w - W \sin \tau,  \nonumber \\
& & \tau' = \tau + \lambda \ln {32 \over \vert w'
\vert} \ \ \ (\mbox{mod } 2 \pi), \label{sm}
\end{eqnarray}

\noindent where $w$ denotes the relative (with respect to the
non-perturbed separatrix value) pendulum energy $w = {H_0 \over
{\cal F}} - 1$, and $\tau$ retains its meaning of the phase angle
of perturbation. The quantities $\lambda$ and $W$ are constant
parameters: $\lambda = \Omega / \omega_0$, where $\omega_0 =
({\cal F G})^{1/2}$ is frequency of small-amplitude phase
oscillations (it is assumed that ${\cal F} > 0$, ${\cal G} >
0$); and

\vspace{-3mm}

\begin{equation}
W = \varepsilon \lambda (A_2(\lambda) + A_2(-\lambda)) = 4
\pi \varepsilon \frac{\lambda^2}{\sinh{\frac{\pi \lambda}{2}}},
\label{W}
\end{equation}

\noindent
where $\varepsilon = a / {\cal F}$, and

\vspace{-3mm}

\begin{equation}
A_2(\lambda) = 4 \pi \lambda {\exp{\pi \lambda \over 2}
\over \sinh (\pi \lambda)}
\label{A2}
\end{equation}

\noindent is a special function, called the Melnikov--Arnold
integral, as defined in~\cite{C79PhR}; expressions for
$A_n(\lambda)$ at any natural $n$ ($n= 1, 2, \dots$) and for
kindred functions are given in~\cite{S00JETP}.

One iteration of map~(\ref{sm}) corresponds to one period of the
model pendulum rotation, or a half-period of its libration. The
motion of system~(\ref{h}) is mapped by Eqs.~(\ref{sm})
asynchronously: the action-like variable $w$ is taken at $\varphi
= \pm \pi$, while the perturbation phase $\tau$ is taken at
$\varphi = 0$. The desynchronization can be removed by a special
procedure \cite{S00JETP}.

An equivalent form of map~(\ref{sm}), as used, e.~g.,
in~\cite{CS84PhyD,S20ASSL}, is given by

\vspace{-3mm}

\begin{eqnarray}
     y' &=& y + \sin x, \nonumber \\
     x' &=& x - \lambda \ln \vert y' \vert + c
                   \ \ \ (\mbox{mod } 2 \pi),
\label{sm1}
\end{eqnarray}

\noindent where $y = {w \over W}$, $x = \tau + \pi$; and

\vspace{-3mm}

\begin{equation}
c = \lambda \ln {32 \over \vert W \vert}
\label{c}
\end{equation}

\noindent is a new constant parameter. This form of the separatrix
map is especially simple-looking; therefore, we call it the ``form
in natural variables.''

On various generalizations of the separatrix map theory (in
particular, for cases of asymmetric perturbation and multiple
interacting resonances) see \cite{S99CM,S20ASSL}.

Linearizing the separatrix map~(\ref{sm1}) in $y$ near a fixed
point (near a resonant $y$ value, and $\sin x$ is not linearized),
one gets the standard map~\cite{C79PhR}:

\vspace{-3mm}

\begin{eqnarray}
     y' &=& y + K \sin x \ \ \ (\mbox{mod } 2 \pi), \nonumber \\
     x' &=& x + y' \ \ \ (\mbox{mod } 2 \pi),
\label{stm}
\end{eqnarray}

\noindent where $K$ is the so-called stochasticity parameter. It
is related to $\lambda$ and the resonant $y=y_\mathrm{res}$ value
by the formula $K = \lambda/y_\mathrm{res}$ \cite{C79PhR}.

Note that $y$ in Eqs.~\ref{stm} is not the same variable as in
Eqs.~(\ref{sm1}). In Eqs.~\ref{stm}, $y$ is a rescaled energy
around a given resonant value.

\section{Lyapunov exponents and dynamical entropy}
\label{sec_de}

The Lyapunov characteristic exponents (LCEs) characterize the rate
of divergence of trajectories close to each other in phase space;
see, e.~g., \cite{LL92}. A nonzero LCE indicates chaotic
character of motion, while the maximum LCE equal to zero is the
signature of regular (periodic or quasi-periodic) motion. The
Lyapunov time (the quantity reciprocal to the maximum LCE)
characterizes predictability time of the motion: in fact, it
provides a lower bound to the predictable dynamics, since the
timescale for any macroscopical change in the orbit is given by
the diffusion time $T_\mathrm{d} \gg T_\mathrm{L}$; see
\cite{CGS22PhyD}. Therefore, calculation of LCEs is one of the
most important instruments in nonlinear dynamics.

Let us consider two trajectories initially close to each other in
phase space. One of them we shall refer to as the {\it guiding}
trajectory and the other as the {\it shadow} one. Let $d(t_0)$ be
the length of the displacement vector directed from the guiding
trajectory to the shadow one at an initial moment~$t = t_0$. The
maximum LCE is defined by the formula~\cite{LL92}:

\begin{equation}
L=\limsup_{{t \to \infty} \atop {d(t_0) \to 0}} {1 \over {t-t_0}}
\ln{d(t) \over d(t_0)} .
\end{equation}

Our numerical data concern the maximum Lyapunov exponent $L$ of
the motion in the chaotic layer in the separatrix map's phase
space, the layer's measure $\mu_\mathrm{ch}$, and the product of
$L$ and $\mu_\mathrm{ch}$. All these data were obtained by means
of numerical experiments with map~(\ref{sm1}).

The initial data were taken inside the chaotic layer. The Lyapunov
exponents were computed by the tangent map method (described,
e.~g., in~\cite{C79PhR}).

In parallel with the maximum Lyapunov exponent $L$ for each
trajectory, the corresponding value of the chaotic layer
half-width $y_\mathrm{b}(\lambda)$ was computed, as attained by
the same trajectory.
The chaotic layer half-width $y_\mathrm{b}$ corresponds to the maximum
deviation of $|y_i|$ (from the unperturbed separatrix) of any
trajectory inside the chaotic layer. It can be obtained as the
$|y_i|$ maximum achieved by a single chaotic trajectory, if the
number of iterations of the map is large enough. The $\lambda$
dependence for $y_\mathrm{b}$ is therefore constructed as follows
\cite{S08PLA}. At each step in $\lambda$, the value of $c$
corresponding to the minimum $y_\mathrm{b}$ is found (over the $0
\leq c < 2 \pi$ interval, by varying the $c$ value with a small
step) and the minimum $y_\mathrm{b}$ is plotted against the
current $\lambda$ value. The obtained $y_\mathrm{b}$ value
corresponds to the case of the least perturbed border, because, by
the given procedure, the contribution of marginal resonances is
minimized.

The number $n_\mathrm{it} = 10^7$ of iterations was found to be
sufficient to saturate the values of both $L$ and $y_\mathrm{b}$,
as verified by comparing the results with those obtained using the
computation time ten times less, i.~e., with $n_\mathrm{it} =
10^6$.

To take into account the ``quasi-sinusoidal'' form of the last
invariant curve that serves as the chaotic layer's border, the
layer half-width value $y_\mathrm{b}$ was calculated as the
average of the maximum and minimum values of $|y|$ at the border,
attained during each computation.

The numerical data cover the interval in $\lambda$ from zero up to
5. This interval is sufficient for our study, because at greater
values of $\lambda$ the asymptotic behaviours of $L$ and $\mu$ are
in principle well established; see \cite{S02CR,S20ASSL} and a
discussion below. The resolution (step) in $\lambda \in [0, 5]$
was set equal to $0.01$. At each step in $\lambda$, the values of
$L$ were computed for 40 values of $c$ equally spaced in the
interval $[0, 2 \pi]$, and the value of $c$ corresponding to the
minimum width of the layer (the case of the least perturbed
border) was fixed. The corresponding values of $L$ and
$y_\mathrm{b}$, as functions of the adiabaticity parameter
$\lambda$, were plotted.

Note that the case of the least perturbed border is generic in
applications, because the border perturbations are strong only
locally in $c$; for details see \cite{S20ASSL}.

\begin{figure}[h!]
\centering
\begin{tabular}{cc}
\includegraphics[width=65mm]{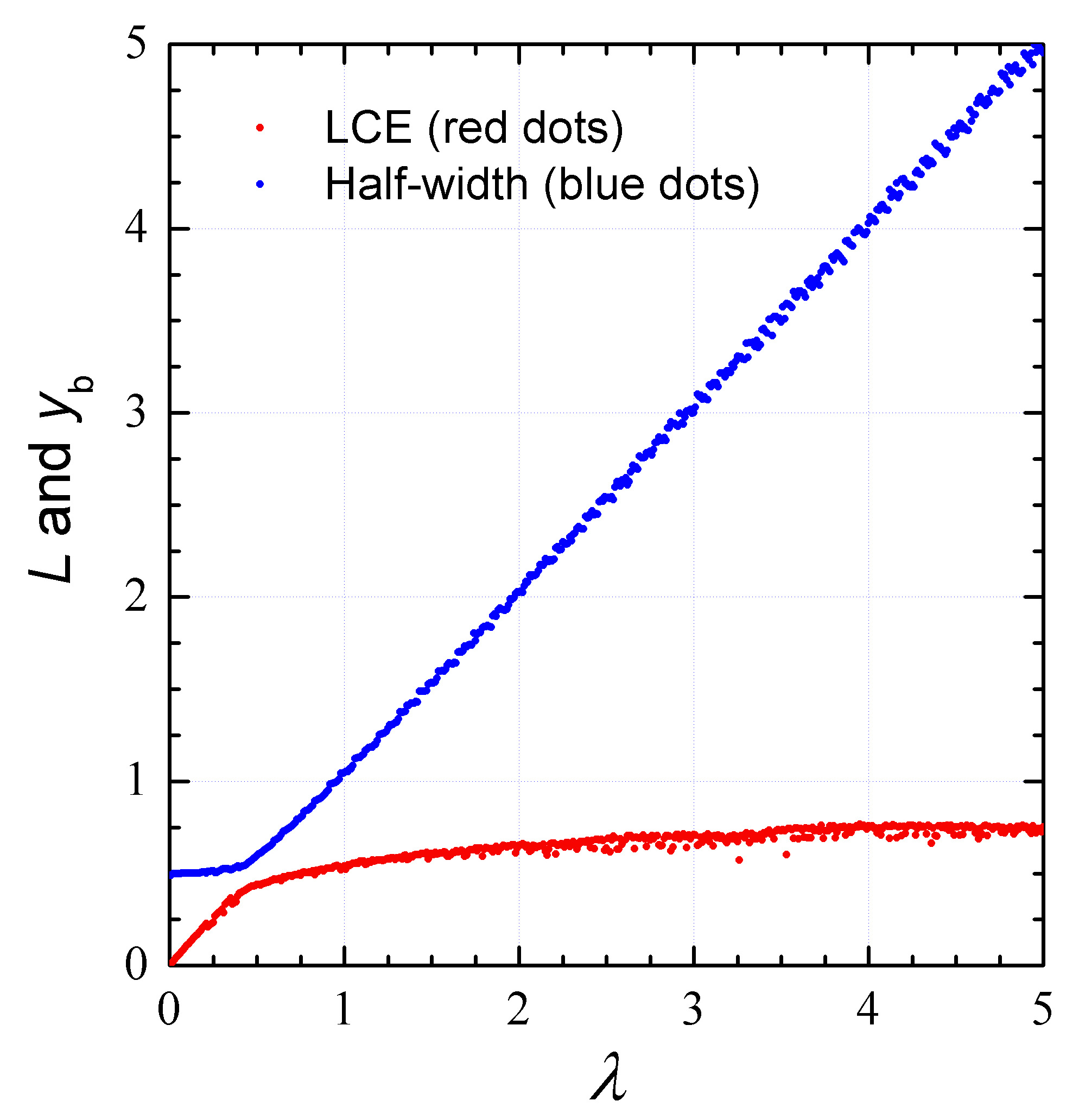}
&
\includegraphics[width=65mm]{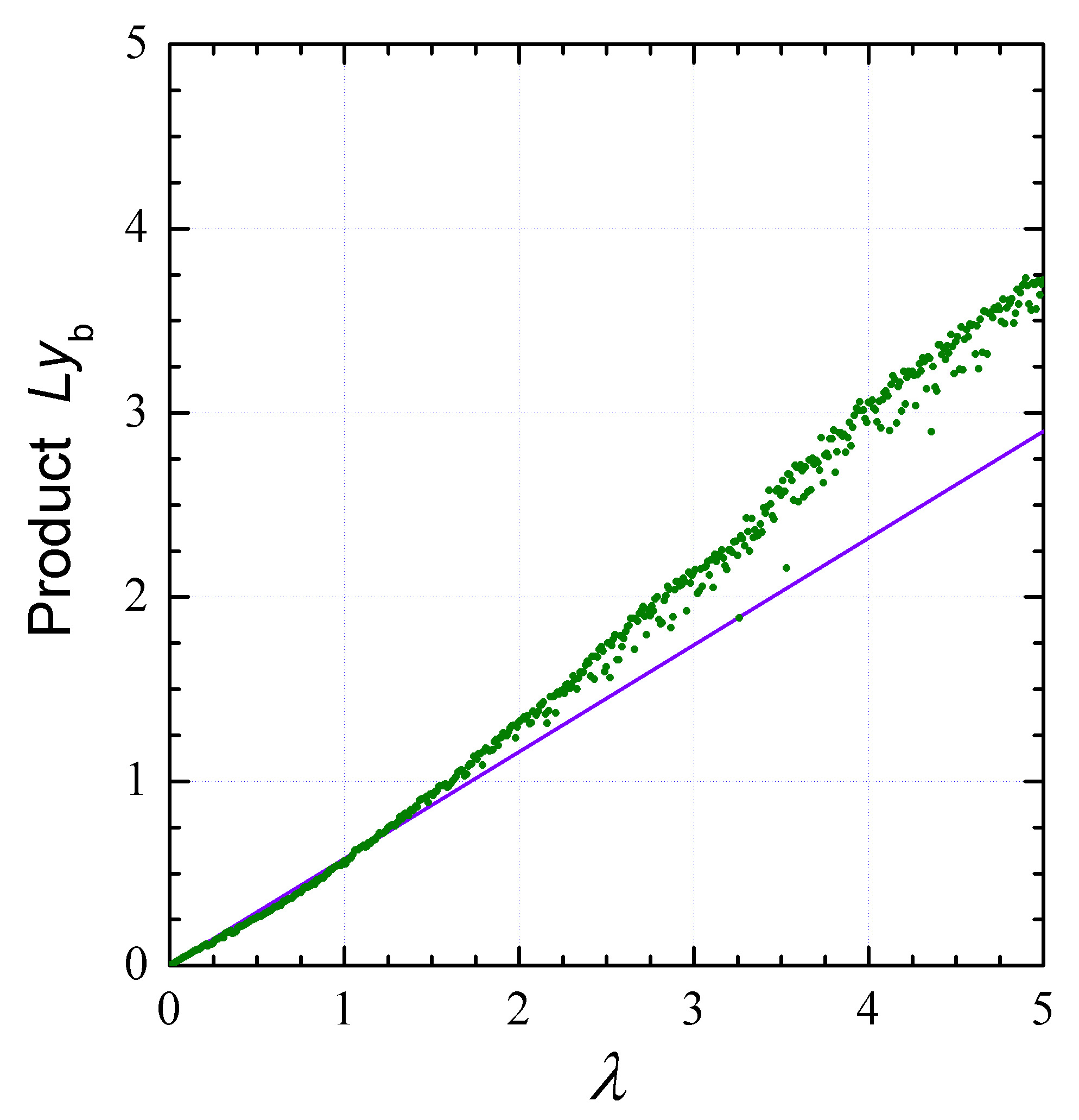}
\end{tabular}
\caption{Left panel: The dependences of the maximum Lyapunov
exponent $L$ (red dots) and the chaotic layer half-width
$y_\mathrm{b}$ (blue dots) on the adiabaticity parameter
$\lambda$, as obtained in our numerical experiments. Right panel:
The $\lambda$ dependence of the dynamical entropy $h = L \cdot
y_\mathrm{b}$ (green dots). The violet straight line represents an
expected function $h(\lambda)$, when the porosity of the chaotic
layer is approximately taken into account, as explained in the
text.} \label{fig1}
\end{figure}

In Fig.~\ref{fig1} (left panel), we present the obtained $\lambda$
dependences for the maximum Lyapunov exponent $L$ (red dots) and
the chaotic layer half-width $y_\mathrm{b}$ (blue dots). The both
dependences do not look quite simple; however, a rational function
fitting is suitable for $L(\lambda)$ \cite{S02CR,S20ASSL}, and a
piecewise-linear fitting for $y_\mathrm{b}(\lambda)$; see
\cite[fig.~5.6]{S20ASSL}. In asymptotic regimes $\lambda \ll 1$
and $\lambda \gg 1$, these dependences allow for theoretical
interpretations; see reviews on them in \cite{S20ASSL}. On the
structure and width parameters of the chaotic layer see also
\cite{CS81,V2004JTePh}.

In Fig.~\ref{fig1} (right panel), the $\lambda$ dependence of the
product $L(\lambda) \cdot y_\mathrm{b}(\lambda)$ is shown (green
dots). The data for $L$ and $y_\mathrm{b}$ presented in the left
panel of Fig.~\ref{fig1} were used to construct it.

Taking into account that $x$ is defined on the segment $[0, 2
\pi]$ and $y_\mathrm{b}$ represents the chaotic layer's
half-width, the phase space chaotic component's measure is given
by the formula

\begin{equation}
\mu_\mathrm{ch}(\lambda) = 4 \pi \sigma(\lambda)
y_\mathrm{b}(\lambda) . \label{mu_ch}
\end{equation}

\noindent Here the coefficient $\sigma$ is the ratio of the
area of the chaotic component inside the chaotic layer to the
layer's total area (as bounded by the layer's external borders);
thus, this coefficient accounts for the layer's ``porosity''
\cite{S04JETPL}. The quantity $1-\sigma$ is nothing but the total
relative area of all regular islands inside the layer.

In the limit $\lambda \gg 1$ and in the least perturbed border
case, the porosity was calculated in \cite{S04JETPL} using the
formula

\begin{equation}
\sigma = \lim_{\lambda \to \infty} y_\mathrm{b}^{-1}
\int\limits_0^{y_\mathrm{b}} \tilde \mu_\mathrm{ch}(y) \, dy =
K_\mathrm{G} \int\limits_{K_\mathrm{G}}^\infty \mu_\mathrm{st}(K)
\, \frac{dK}{K^2} , \label{sg}
\end{equation}

\noindent where $\tilde \mu_\mathrm{ch}(y)$ is the local relative
measure of the chaotic component of the separatrix map phase
space, and $\mu_\mathrm{st}(K)$ is the relative measure of the
chaotic component of phase space of the standard map~(\ref{stm}).
The quantity $K_\mathrm{G}$ is the critical value of the
stochasticity parameter $K$ \cite{M92RvMP}: $K_\mathrm{G} =
0.971635406\ldots$.

According to \cite{S04JETPL}, in the limit $\lambda \to \infty$
one has $\sigma \approx 0.780$, and, in the function

$$ h(\lambda) = C_1 \lambda , $$

\noindent the coefficient

$$ C_1 = C_\mathrm{h} \sigma \approx 0.625 , $$

\noindent where $C_\mathrm{h} \approx 0.801$ is Chirikov's
constant (the maximum Lyapunov exponent of the separatrix map
in the limit $\lambda \to \infty$, see \cite{S04JETPL}).

The green dots in the right panel of Fig.~\ref{fig1} represent the
multiplication of the two curves in the left panel. Although the
slope for $y_\mathrm{b}(\lambda)$ at $\lambda > 1/2$ (left panel,
blue dots) seems to be close to strictly 1, a closer inspection of
the graph shows that, at $\lambda$ near $1/2$, the slope deviates
from unity quite essentially. The visible restricted deviations of
the dynamical entropy (right panel, green dots), as a function of
$\lambda$, from the linear law might be due to numerical
indefiniteness of the layer porosity, which should be taken into
account when calculating the chaotic domain measure, but which is
indeed hard to evaluate.

In Fig.~\ref{fig1} (right panel), the violet straight line
hypothetically describes the expected dependence $h(\lambda)$, on
taking into account the chaotic layer porosity $\sigma$ (in the
limit $\lambda \gg 1$, where $\sigma \approx 0.78$). Upon this
improvement, the over-all slope of the dependence obviously looks
much more close to 1/2 (the initial slope observed at $\lambda
\lesssim 1/2$).

Form~(\ref{sm1}) of the separatrix map is that in natural
variables ($x$, $y$), as defined above. We see that, adopting this
form, one has

\begin{equation}
h \approx 2 \pi \lambda , \label{h_yx}
\end{equation}

\noindent because the layer's span in $x$ is $2 \pi$.

Returning to the separatrix map form~(\ref{sm}), i.~e., the form
in original variables ($\tau$, $w$), in which the layer half-width
$w_\mathrm{b} = W y_\mathrm{b}$, and $W$ is given by
formula~(\ref{W}), one finds the dynamical entropy

\begin{equation}
h \approx 2 \pi \lambda W = 8 \pi^2 \varepsilon
\frac{\lambda^3}{\sinh{\frac{\pi \lambda}{2}}} \label{h_wtau}
\end{equation}

\noindent over the whole range $\lambda \in [ 0, \infty ]$. Note
that $h(\lambda)$ turns out to be expressed through the
``Bose--Einstein function'' $1/\sinh(\alpha \lambda)$, where
$\alpha$ is constant.

If, in the initial Hamiltonian~(\ref{h}) one sets $k=1/2$ instead
of $k=1$, keeping $a=b$, then, according to \cite[eqs.~(12) and
(A.12)]{S00JETP},

\begin{equation}
W = \varepsilon \lambda (A_1(\lambda) + A_1(-\lambda)) = \frac{2
\pi \varepsilon \lambda}{\cosh{\frac{\pi \lambda}{2}}} ,
\label{W12}
\end{equation}

\noindent where $\varepsilon = a / {\cal F}$, as defined above.
Therefore, for the separatrix map in original variables ($\tau$,
$w$) one gets the dynamical entropy in the form

\begin{equation}
h \approx 2 \pi \lambda W = 4 \pi^2 \varepsilon
\frac{\lambda^2}{\cosh{\frac{\pi \lambda}{2}}} , \label{h_wtau12}
\end{equation}

\noindent again over the whole range $\lambda \in [ 0, \infty ]$.
Note that here $h(\lambda)$ is expressed through the
``Fermi--Dirac function'' $1/\cosh(\alpha \lambda)$.

The examples, given by Eqs.~(\ref{h_wtau}) and (\ref{h_wtau12}),
demonstrate that the intrinsic quasilinear (or perhaps exactly
linear) relationship $\lambda$--$h$ for the separatrix map in
natural variables forms a basis for calculating the dynamical
entropy for any perturbed nonlinear resonance, in its first
fundamental model, if Melnikov--Arnold integrals are known.

\section{Discussion}
\label{sec_discussion}

The adiabaticity parameter $\lambda$, which is the ratio of the
perturbation frequency $\Omega$ to the frequency $\omega_0$ of
small-amplitude oscillations on resonance, is the main parameter
of the problem. It characterizes the separation of the perturbing
and guiding resonances in units of one quarter of the guiding
resonance width. Indeed, $\lambda = \Omega / \omega_0$, and the
separation of resonances in frequency space is equal to $\Omega$,
whereas the guiding resonance width is equal to $4 \omega_0$
\cite{C79PhR}. If, in the Hamiltonian system~(\ref{h}), the
perturbation frequency is relatively large, the separation of
resonances in the canonical momentum is also large and they almost
do not interact. On reducing the frequency of perturbation, they
approach each other, and appreciable chaotic layers emerge in the
vicinity of the separatrices. On reducing the frequency of
perturbation further on, the layers merge into a single chaotic
layer, due to the strong resonance overlap. Therefore, the
adiabaticity parameter $\lambda$ can be regarded as a kind of a
resonance-overlap parameter. In the asymptotic limit of the
adiabatic perturbation, $\lambda \ll 1$, the resonances in the
multiplet strongly overlap, while in the asymptotic limit of the
non-adiabatic perturbation, $\lambda \gg 1$, the resonances are
separated and do not interact.

Up to here we have considered the motion inside the main chaotic
layer of the separatrix map, ignoring the smaller chaotic layers
not connected heteroclinically with the main layer; these layers
are present around the resonances external to the main layer. How
large can be the contribution of these external chaotic domains to
the total dynamical entropy of the separatrix map, i.e., the
dynamical entropy calculated over the whole phase space of the
map? It turns out to be rather small, as we find in the following.

According to \cite[eq.~10]{S04JETPL}, at $\lambda \gg 1$, the
dynamical entropy inside the main chaotic layer can be represented
as

\begin{equation}
h(\lambda) =
C_1 \lambda , \label{hK1}
\end{equation}

\noindent with
\begin{equation}
C_1 = K_\mathrm{G} \int_{K_\mathrm{G}}^\infty
h_\mathrm{standard}(K) \frac{\mathrm{d}K}{K^2} \approx 0.625 .
\label{C1}
\end{equation}

\noindent Here $h_\mathrm{standard}(K)$ is the dynamical entropy
of the standard map as a function of the stochasticity parameter
$K$.

For the dynamical entropy {\it outside} the main layer,
one analogously arrives at

\begin{equation}
h_\mathrm{aux}(\lambda) = C_2 \lambda , \label{hK2}
\end{equation}

\noindent where
\begin{equation}
C_2 = K_\mathrm{G} \int_0^{K_\mathrm{G}} h_\mathrm{standard}(K)
\frac{\mathrm{d}K}{K^2} . \label{C2a}
\end{equation}

\noindent A theoretical approximation for the dynamical entropy
$h_\mathrm{standard}(K)$ of the standard map at $K \in [0, 1]$ is
given by the function \cite{S04PLA}:

\begin{equation}
h_\mathrm{standard}(K) = A K^{1/2} \exp \left(
-\frac{\pi^2}{K^{1/2}} \right) \label{hexp}
\end{equation}

\noindent with $A = 929.6 \pm 4.0$. Then, one obtains

\begin{equation}
C_2 = A K_\mathrm{G} \int_0^{K_\mathrm{G}} \exp \left(
-\frac{\pi^2}{K^{1/2}} \right) \frac{\mathrm{d}K}{K^{3/2}} =
\frac{2 A K_\mathrm{G}}{\pi^2} \exp (x) \vert_{x=-\infty}^{x=\pi^2
/ \sqrt{K_\mathrm{G}}} \approx 0.00820 .
\label{C2}
\end{equation}

\noindent The contribution of the external chaotic component to
the total dynamical entropy of the separatrix map is thus
relatively small, because $C_2 < 0.01$. Indeed, in the given limit
of large $\lambda \gg 1$, the calculated total dynamical entropy
$h_\mathrm{total} = C_1 \lambda + C_2 \lambda \approx 0.633
\lambda$ is almost the same as the dynamical entropy $h = C_1
\lambda \approx 0.625 \lambda$ calculated for the motion in the
main chaotic layer.

\begin{figure}[h!]
\centering
\begin{tabular}{cc}
\includegraphics[width=65mm]{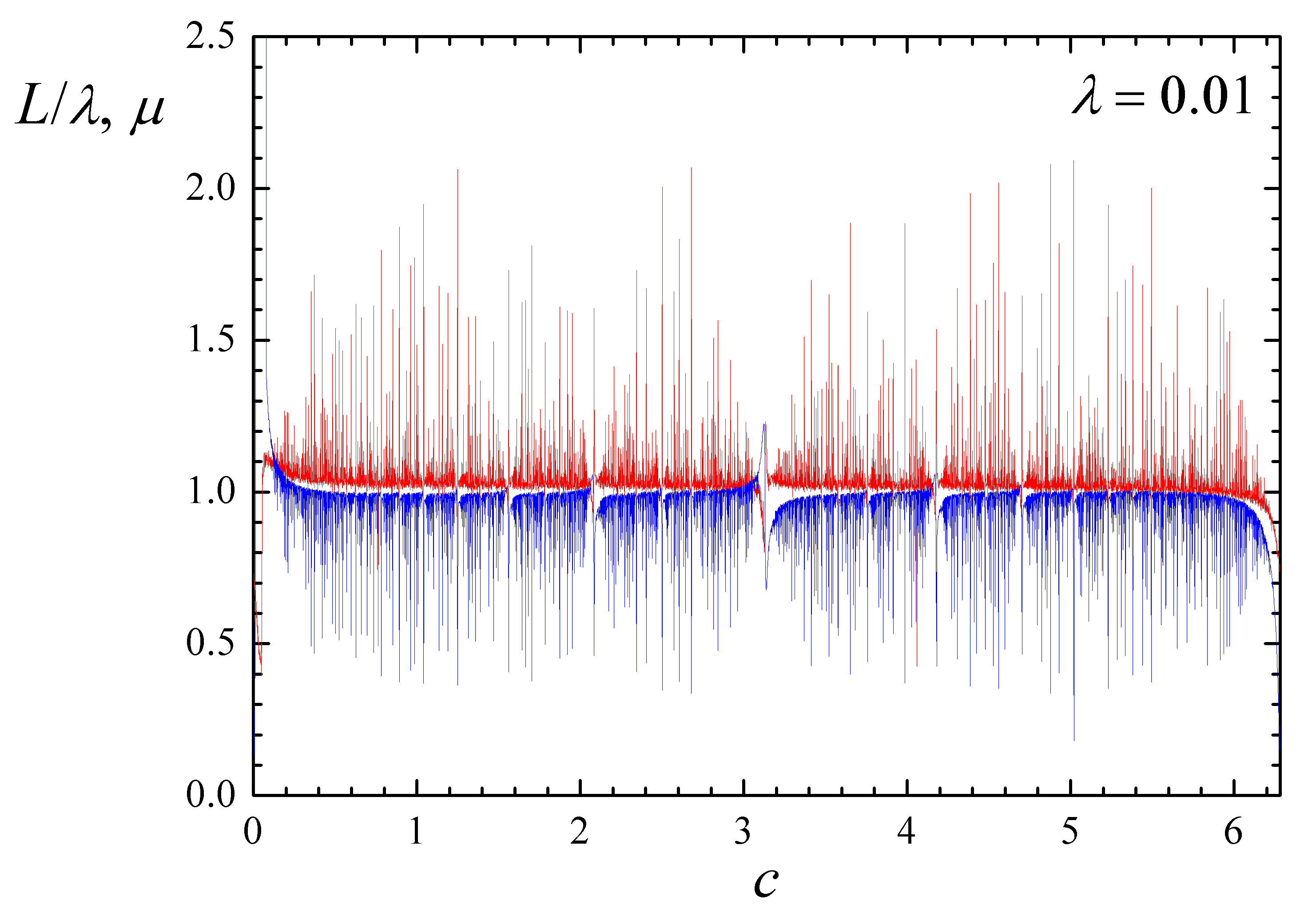}
&
\includegraphics[width=65mm]{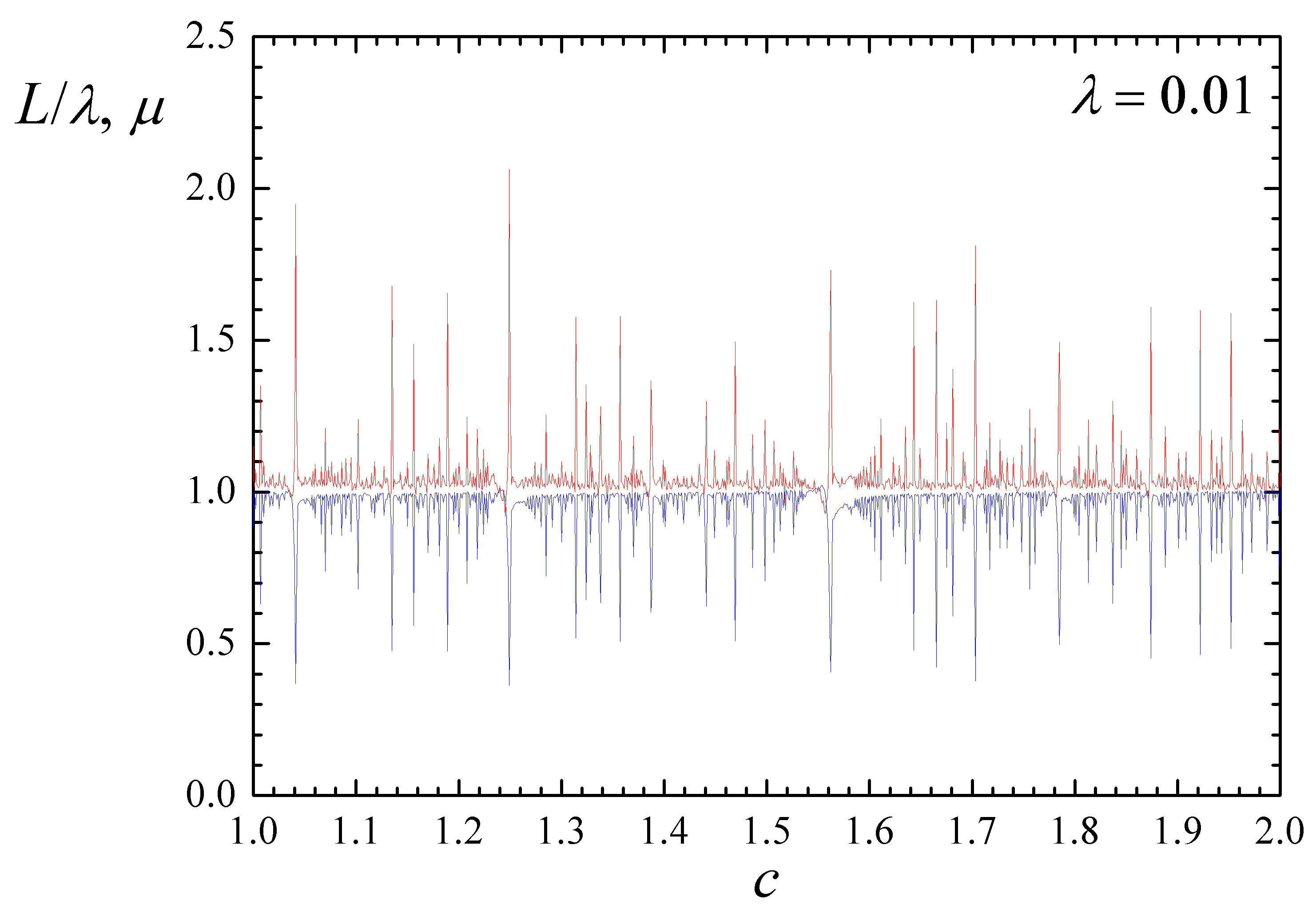}
\end{tabular}
\caption{Left panel: The parameter $c$ dependences of $L /
\lambda$ (the $\lambda$-normalized maximum Lyapunov exponent $L$;
red dots) and the chaotic domain measure $\mu$ (blue dots), at a
fixed value of $\lambda$ (set equal to 0.01). Right panel: the
graph's detail given in a higher resolution.} \label{fig2}
\end{figure}

Let us briefly discuss the role of the parameter $c$. By
considering the dynamical entropy instead of the Lyapunov
exponents and the chaotic domain measure, the sharp local
variations of the latter two quantities, prominent in their
dependences on $\lambda$, are also expected to be smoothed out
when one considers dependences on the parameter $c$. This is
illustrated in Fig.~\ref{fig2}, where we present computed
dependences of the $\lambda$-normalized maximum Lyapunov exponent
$L$ and the chaotic domain measure $\mu$ on the parameter $c$ at a
fixed value of $\lambda$ ($\lambda = 0.01$). The measure $\mu$ was
computed here following an approach adopted in
\cite{C79PhR,S04PLA,S08MN}; it takes into account the variation of
the chaotic layer porosity (the presence of regular islands inside
the layer) with $c$. Namely, to compute $\mu$, the traditional
``one trajectory method'' was used: the number of cells, explored
by a single trajectory on a grid exposed over the chaotic domain,
was calculated. The grid was set to consist of $2000 \times 2000$
pixels. At each step in $c$, the map~(\ref{sm1}) was iterated
$10^8$ times, sufficient to saturate the computed values of $\mu$
and $L$. The step in $c$ is was set to $0.001$.

From the graphs in Fig.~\ref{fig2}, it is obvious that the
variations in $L$ and $\mu$ are anticorrelated, and, therefore,
the multiplication of the two curves, which gives the dynamical
entropy, provides a significantly smoothed curve. This again
underlines that the dynamical entropy is a much more fundamental
property, as compared to $L$ and $\mu$. This example demonstrates
that the $\lambda$-normalized dynamical entropy $h$ might stay
constant (at least approximately) not only with variation of
$\lambda$, but also with variation of $c$.

The numerical methods used above require quite a lot of computing
time to achieve reliable results. In this respect, it is worth
noting that both the maximum Lyapunov exponent and the chaotic
layer width may also be conveniently computed by applying
techniques of the dynamical exponent curves
\cite{Gao99PRL,Gao06PRE}, which have straightforward
interpretations for these computable quantities,
and may require fewer data points than used above. In future,
efforts can be made to connect further relevant studies with the
exponent curves.

Concerning further applications of the results presented in this
article, one should outline that, if the constancy of the
$\lambda$-normalized dynamical entropy were verified analytically,
this constancy would provide a convenient framework to evaluate
the measure $\mu$ of chaotic domains in phase space in various
physical models (mentioned already above), as soon as the maximum
Lyapunov exponent is calculated. This is important because any
calculations of $\mu$ are much more complicated and
computationally expensive that those of $L$.

As already mentioned in the Introduction, the employed model
relates with particularly interesting physical phenomena and is
both applicable to a broad spectrum of physical phenomena. It
should also be noted that the given study subject can be expanded,
to encompass the separatrix map generalizations. The generalized
separatrix maps  are characterized by the $\tau$ increments (in
equations~(\ref{sm})) that are algebraic (power-law with various
power-law indices), not logarithmic ones; see \cite{S20ASSL} for
examples. A particular generalized map is used to describe
dynamical environments of cometary nuclei \cite{LSR18Ic}.

\section{Conclusions}

In this article, we have calculated the maximum Lyapunov exponent
of the motion in the separatrix map's chaotic layer, along with
calculation of its width, as functions of the adiabaticity
parameter $\lambda$. The case of the layer's least perturbed
border has been considered.

However, although these obtained dependences are not at all
simple-looking, the dynamical entropy $h$ of the separatrix map
(in natural variables) has turned out to be an almost-linear
function of $\lambda$; in other words, if normalized by $\lambda$,
the entropy is a quasi-constant.

Thus, the formula for dynamical entropy of the separatrix
map~(\ref{sm1}) has turned out to be rather simple (quasilinear in
$\lambda$), in contrast to the much more complicated behavior of
the maximum Lyapunov exponent and the layer width. This is no
wonder, since the dynamical entropy is a more fundamental property
of the motion than the latter two.

To calculate the dynamical entropy more precisely, the layer's
porosity $\sigma$ must be taken into account as a function of
$\lambda$; this represents a perspective numerical-experimental
and/or theoretical problem. As demonstrated in our study, even an
approximate accounting for $\sigma$ makes the observed
numerical-experimental dependence $h(\lambda)$ much closer to a
linear one.

The intrinsic quasilinear (or perhaps exactly linear) relationship
$\lambda$--$h$ for the separatrix map in natural variables forms a
basis for calculating the dynamical entropy for any perturbed
nonlinear resonance in the first fundamental model, as soon as the
corresponding Melnikov--Arnold integral is known.

%\section*{Data Availability Statement}
%This publication is theoretical work that does not require supporting research data.

\section*{Acknowledgements}
The author is most thankful to Pablo Cincotta for invaluable comments.
The author would like to thank the referees for
remarks and comments which greatly improved the manuscript.

\medskip

\noindent The author declares no conflict of interest.

\newpage

\newpage

\end{document}